\documentclass[fleqn,twoside]{article}
\usepackage{espcrc2}

\usepackage{graphicx}

\newcommand{\AmS}{{\protect\the\textfont2
  A\kern-.1667em\lower.5ex\hbox{M}\kern-.125emS}}

\def\simgt{\,\rlap{\lower 3.5 pt\hbox{$\mathchar \sim$}}\raise 1pt \hbox {$>$}\,}
\def\simlt{\,\rlap{\lower 3.5 pt\hbox{$\mathchar \sim$}}\raise 1pt \hbox {$<$}\,}

\title{
Fluctuations in the vicinity of the phase transition line 
for two flavor QCD
\thanks{Presented by S. Ejiri. This work is supported by 
BMBF grant No.06BI102, DFG grant KA 1198/6-4, 
PPARC grant PPA/a/s/1999/00026 and KBN grant 2P03 (06925).
}}

\author{ S.~Ejiri\rlap,
\address{Fakult\"{a}t f\"{u}r Physik, Universit\"{a}t 
Bielefeld, D-33615 Bielefeld, Germany}
C.R.~Allton\rlap,\address{Department of Physics, University of 
Wales Swansea, Singleton Park, Swansea, SA2 8PP, U.K.} 
M.~D\"{o}ring\rlap,$^{\rm a}$ S.J.~Hands\rlap,$^{\rm b}$ 
O.~Kaczmarek\rlap,$^{\rm a}$ F.~Karsch\rlap,$^{\rm a}$ 
E.~Laermann\rlap,$^{\rm a}$ and K.~Redlich\rlap,$^{\rm a}$ \hspace{-2mm}
\address{Institute of Theoretical Physics, University of Wroclaw,
PL-50204 Wroclaw, Poland} }

\begin{document}

\begin{abstract}
We study the susceptibilities of quark number, isospin number 
and electric charge in numerical simulations of lattice QCD at 
high temperature and density. 
We discuss the equation of state for 2 flavor QCD at non-zero 
temperature and density. Derivatives of $\ln Z$ with respect to quark 
chemical potential $(\mu_q)$ are calculated up to sixth order. 
From this Taylor series, the susceptibilities are estimated as functions 
of temperature and $\mu_q$. 
Moreover, we comment on the hadron resonance gas model, which explains 
well our simulation results below $T_c$.
\vspace{1pc}
\end{abstract}

\maketitle

Thermal fluctuations near the critical temperature $(T_c)$ provide 
important information for the understanding of the properties of the 
QCD phase transition at high temperature and density. 
Baryon number fluctuations are expected to diverge at 
the endpoint of the first order chiral phase transition line. 
The electric charge fluctuation in heavy-ion collisions is 
one of the most promising experimental observables to identify 
the critical endpoint.
The fluctuations are estimated by the susceptibilities of quark number  
$(\chi_q)$ and isospin number $(\chi_I)$, which are given by 
$\chi_q= \partial^2 p / \partial \mu_q^2$ and 
$\chi_I= \partial^2 p / \partial \mu_I^2$, 
where $\mu_q=(\mu_u+\mu_d)/2$, $\mu_I=(\mu_u-\mu_d)/2$.  
$\mu_{u[d]}$ is the chemical potential for the $u[d]$ quark.
The charge fluctuation is $\chi_{C} = \chi_q/36 + \chi_I/4$ 
for $\mu_u=\mu_d$.

The simulation of QCD at non-zero $\mu_q$ is known to be 
difficult. However, studies based on a Taylor expansion with respect to 
$\mu_q$ at $\mu_q=0$ turned out to be an efficient technique to 
investigate the low density regime, interesting for heavy-ion collisions. 
We discussed in \cite{us02} the relation between the line of 
constant pressure (energy density) and the phase transition line 
calculating the second derivative of pressure. 
From the study of the fourth derivatives \cite{us03}, we observed 
strong enhancement of $\chi_q$ near $T_c$, suggesting the presence of 
the critical point at finite $\mu_q$. 
In this study, we take the calculation of $p$ to $O(\mu_q^6)$, and 
that of $\chi_q$ $[\chi_I]$ to $O(\mu_q^4)$. 
This enables us to estimate the shift of the peak position of 
$\chi_q$ in the $(T,\mu_q)$ plane. 
Also, by calculating the ratio of the Taylor expansion coefficients, 
the application range of the hadron resonance gas model \cite{KRT} 
is discussed.

\begin{figure}[t]
\centerline{
\includegraphics[width=6.1cm]{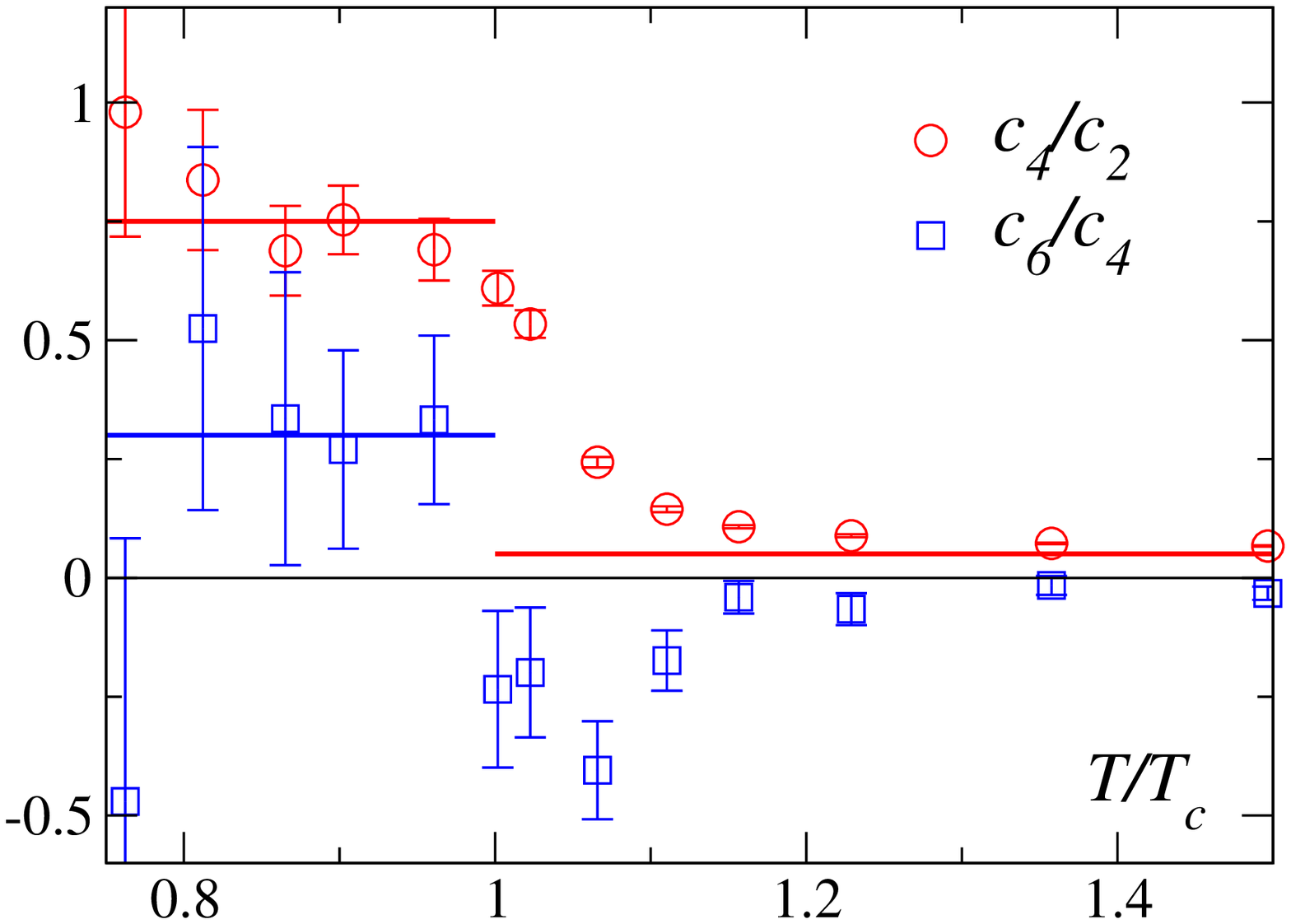}
}
\centerline{
\includegraphics[width=6.1cm]{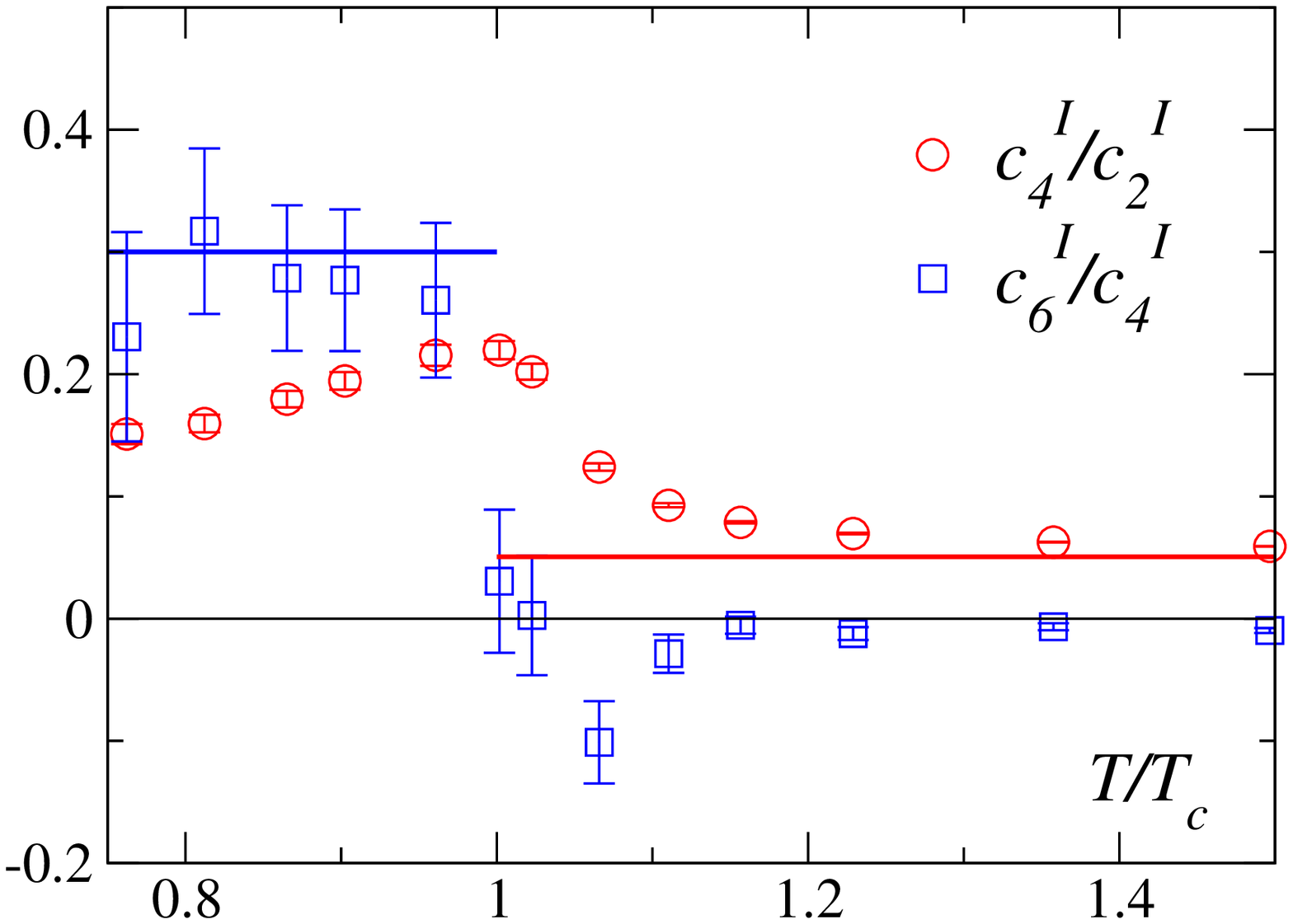}
}
\vspace*{-9mm}
\caption{
The ratio of Taylor expansion coefficients for $\chi_q$ (upper) 
and $\chi_I$ (lower).
}
\vspace*{-4mm}
\label{fig:c2c4c6}
\end{figure}

\paragraph{Quark gluon gas, hadron resonance gas}
We define the expansion coefficients 
$c_n$ and $c_n^I$ by 
$p/T^4 =(\ln Z)/(VT^3) 
\equiv \sum_{n=0}^{\infty} c_n (\mu_q/T)^n$ and 
$\chi_q [\chi_I]/T^2 \equiv 
\sum_{n=2}^{\infty} n(n-1) c_n [c_n^I] (\mu_q/T)^{n-2}$ 
for $\mu_u=\mu_d$.
We expect the equation of state to approach that of a free quark-gluon 
gas (Stefan-Boltzmann (SB) gas) in the high temperature limit. 
The coefficients in the SB limit for $N_{\rm f}=2$, $\mu_I=0$ 
are well known as 
$c_2=c_2^I=1, c_4=c_4^I =1/(2\pi^2)$ and $c_n=c_n^I=0$ for $n \geq 6$. 

On the other hand, in the low temperature phase QCD is well modelled 
by a hadron resonance gas. 
If the interaction between these hadrons can be neglected, 
the pressure is obtained by summing over the contributions from 
all resonance states of hadrons.
The contribution to $p/T^4$ from a particle which has mass $m_i$, 
baryon number $B_i$ and third component of isospin number $I_{3i}$ is given by 
\begin{eqnarray}
\frac{p_{m_i}}{T^4}=\frac{1}{2 \pi^2} \left( \frac{m_i}{T} \right)^2 
\sum_{l=1}^{\infty}{\frac{\eta^{l+1}}{l^{2}} 
K_2 \left( \frac{lm_i}{T} \right) z^l} , 
\label{eq:frg}
\end{eqnarray}
where $z=\exp[(3B_i \mu_q + 2I_{3i} \mu_I)/T]$, $K_2$ is the modified 
Bessel function, and $\eta$ is $1$ for mesons and $-1$ for baryons.
Moreover, since $m_i/T \gg 1$ for all baryons and 
$K_2(x) \approx \sqrt{\pi/2x} \exp(-x)$, 
the first term of $l$ is dominant in the baryon sector.
Therefore, the pressure can be written as 
$p/T^4=G(T)+F(T) \cosh (3 \mu_q/T)$ for $\mu_I=0$, where 
$F(T)$ is the baryonic component of $p/T^4$ at $\mu_q=0$
and $G(T)=G_{\rm sing}(T)+G_{\rm trip}(T)$ is the mesonic part 
which has isosinglet and isotriplet components \cite{KRT}. 
Similarly, we obtain $\chi_q/T^2=9F(T) \cosh (3 \mu_q/T)$ and 
$\chi_I/T^2=G^I(T)+F^I(T) \cosh (3 \mu_q/T)$. 
Here, the mesonic component for $\chi_q$ is zero because mesons 
$(B_i=0)$ are independent of $\mu_q$. 
Therefore, $c_4/c_2=3/4$, $c_6/c_4=3/10$ and $c_6^I/c_4^I=3/10$ in the 
region where the non-interacting hadron 
resonance gas provides a good approximation. 

We investigate these coefficients. 
Simulations are performed on a $16^3 \times 4$ lattice for 
2 flavor QCD. The $p4$ improved action \cite{HKS} is employed at 
$ma=0.1$, which gives $m_{PS}/m_{V} = 0.7$. 
The number of configurations is 1000-5000 for each $T$.
We use the random noise method \cite{us02,us03}. 
The results for $c_{n+2}/c_n$ and $c_{n+2}^I/c_n^I$ 
are shown in Fig.~\ref{fig:c2c4c6}. 
We find that these results are consistent with the prediction from 
the hadron resonance gas model for $T/T_c \leq 0.96$ and
approach the SB values, 
i.e. $c_4/c_2= c_4^I/c_2^I=1/2\pi^2$, $c_6/c_4=c_6^I/c_4^I=0$, 
in the high temperature limit. 
These results suggest that the models of free quark-gluon gas
and hadron gas seem to explain the behavior of thermodynamical 
quantities well except in the narrow regime near $T_c$.

\begin{figure}[t]
\centerline{
\includegraphics[width=6.1cm]{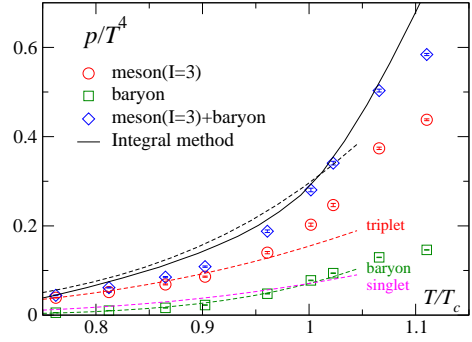}
}
\vspace*{-9mm}
\caption{
Pressure at $\mu_q=0$ as a function of $T$.
}
\vspace*{-4mm}
\label{fig:pres}
\end{figure}

\paragraph{Mesonic and baryonic contributions}
Moreover, if the data is consistent with the hadron resonance gas model, 
we can separate $\chi_I$ into the contributions from isotriplet meson 
and baryons, since $G^I+F^I=2c_2^I$ and $9F^I/2=12c_4^I$. 
Isosinglet mesons do not contribute to $\chi_I$ because $I_{3i}=0$. 
Also, $9F=2c_2$ from $\chi_q.$ 
Furthermore, for the case of $m_{\pi} \gg T$, the first term $(l=1)$ 
in Eq.(\ref{eq:frg}) dominates; 
this approximation holds for our present simulations where 
$m_{\pi}/T \simeq 4$-$5$ but will no longer be valid with physical 
quark mass values. In this approximation,
the ratio of $\partial^2 (p_{mi}/T^4)/\partial (\mu_I/T)^2$ and $p_{mi}/T^4$ 
equals $(2I_{3i})^2$ for $I_{3i}=\{-1,0,1\}$. 
We thus find for the contribution of the triplet meson 
part, $G_{\rm trip}=3G^I/8$.
We then can estimate the contribution of the 
isosinglet mesons to $p/T^4,$ $G_{\rm sing},$ by comparing 
$F+G_{\rm trip}$ with the total pressure. 
In Fig.~\ref{fig:pres}, we plot the baryon component $F=2c_2/9$ (square), 
the triplet meson component $G_{\rm trip}=3c_2^I/4-c_4^I$ (circle) and 
the total (diamond). The solid line is the pressure obtained with 
the integral method in \cite{KLP}. 
The dashed lines in Fig.~\ref{fig:pres} represent the predictions from 
the hadron resonance gas model for the mass parameter 
of our simulations. The resonance states for this quark mass are 
adjusted by the method described in \cite{KRT}.
The simulation results for $p/T^4$ obtained with the integral method, 
together with the triplet meson and baryon contributions, are 
surprisingly well reproduced by this model calculation. 
The result for the singlet meson contribution may explains the 
difference in pressure obtained from the integral method and that 
from the Taylor expansion which neglects the singlet part.

\paragraph{Susceptibilities at $\mu_q \neq 0$} 

\begin{figure}[t]
\centerline{
\includegraphics[width=6.1cm]{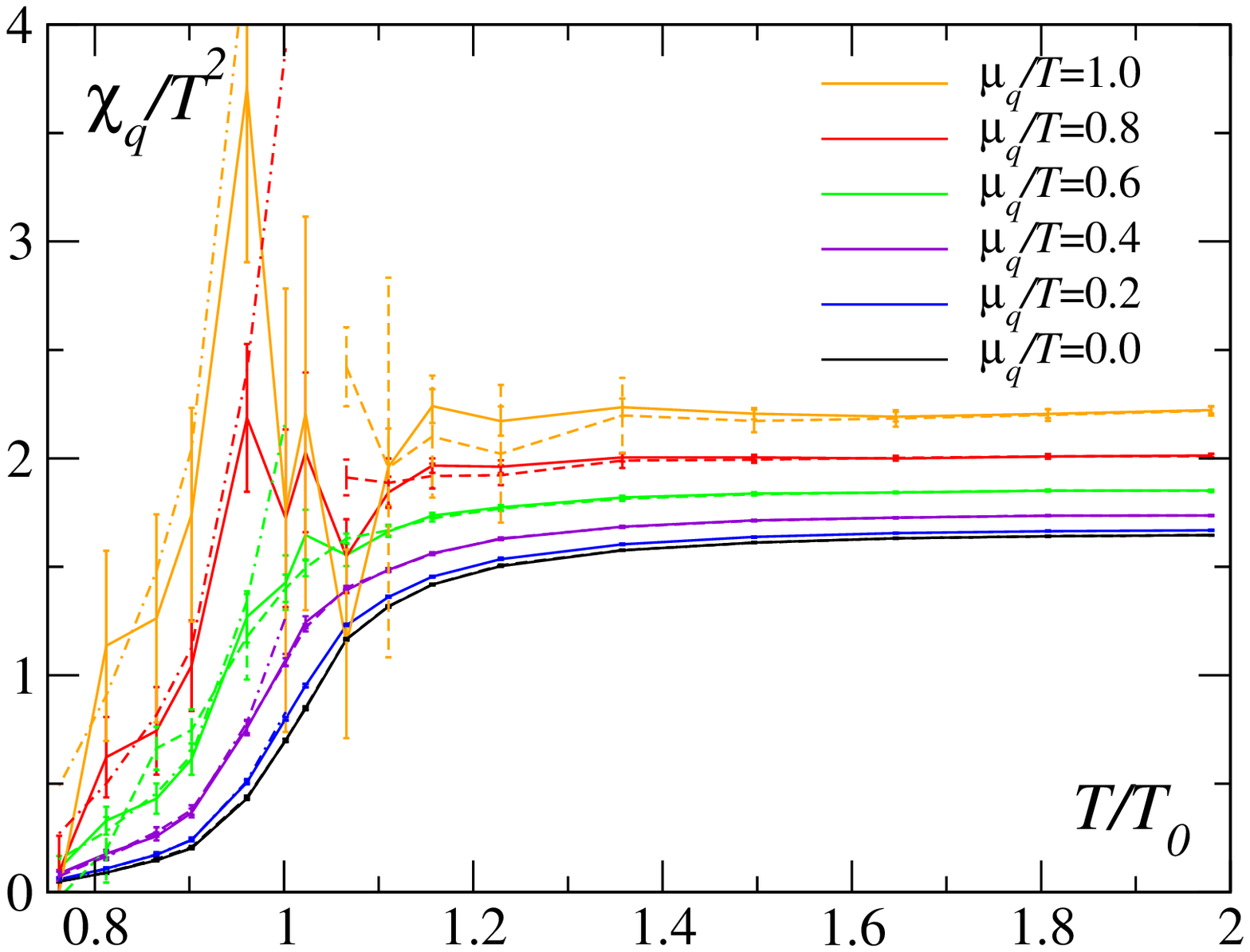}
}
\centerline{
\includegraphics[width=6.1cm]{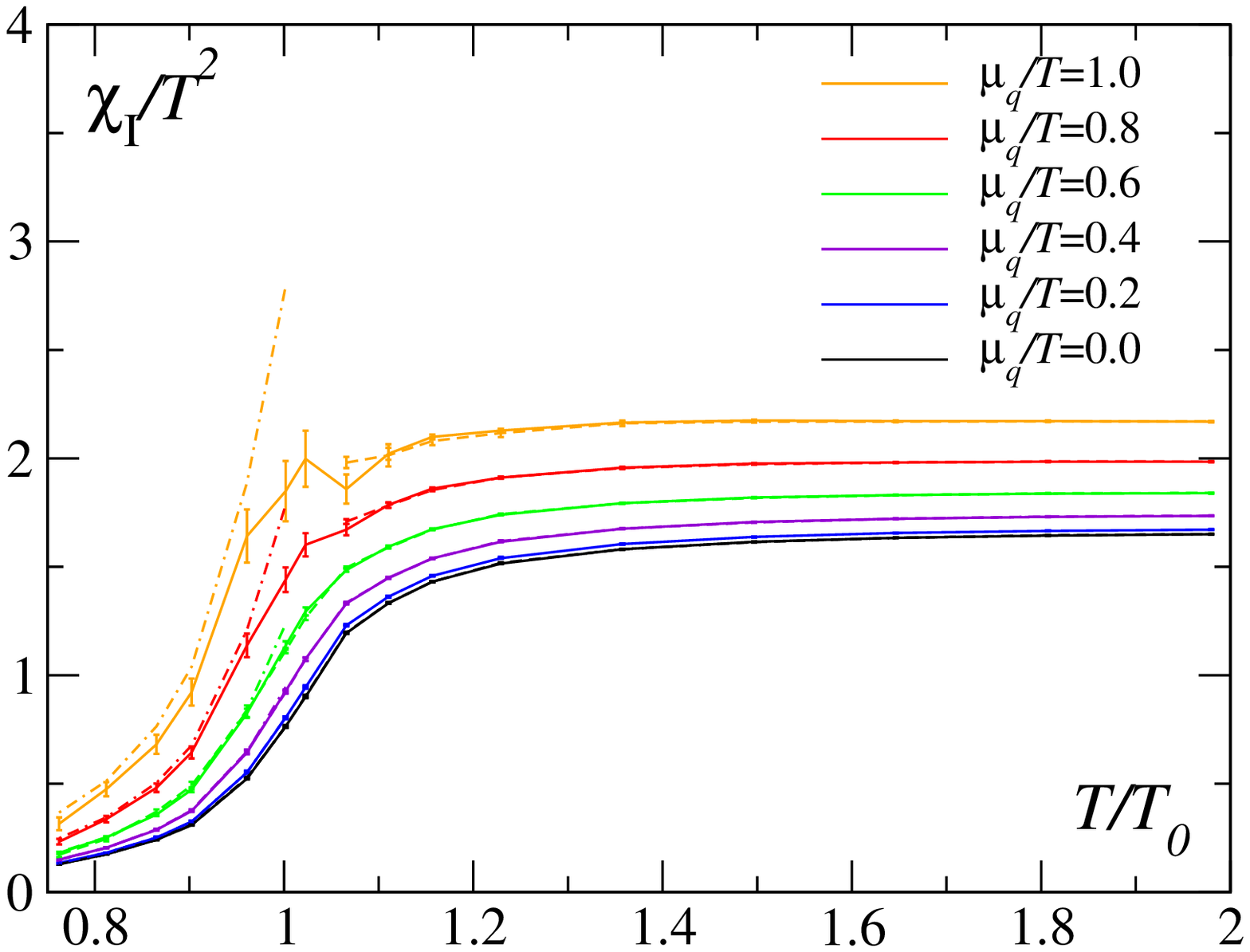}
}
\vspace*{-9mm}
\caption{
Quark number (upper) and isovector (lower) susceptibilities as 
a function of $T$ for each fixed $\mu_q/T$. 
$T_0$ is $T_c$ at $\mu_q=0$.
}
\vspace*{-4mm}
\label{fig:nsus}
\end{figure}

Next, we calculate quark number and isovector susceptibilities, 
using three methods, in a range of $0 \leq \mu_q/T \leq 1$. 
The data connected by solid lines in Fig.~\ref{fig:nsus} are obtained by 
$\chi_q/T^2 = 2c_2 + 12c_4(\mu_q/T)^2 + 30c_6(\mu_q/T)^4$ 
and the corresponding equation for $\chi_I$. 
Dot-dashed lines are from the hadron resonance gas model, using $F, F^I$ 
and $G^I$. This turned out to be a good approximation 
for $T/T_c \simlt 0.96$. 
Dashed lines are calculated by the reweighting method with an 
approximation in \cite{us02}. Here the truncation error is 
$O(\mu_q^6)$ but the effect from higher order terms of $\mu_q$ is 
partially included. The sign problem in the reweighting method 
is serious when the complex phase fluctuations of the quark 
determinant are large at large $\mu_q/T$. 
We omitted data for which the standard deviation of the complex phase 
is larger than $\pi/2$. The difference among the three results is caused by 
the approximation in the higher order terms of $\mu_q/T$. 

Since the statistical error of $c_6$ is still large near $T_c$, the peak of 
$\chi_q$ seems to lose its statistical significance. 
However, as seen in Fig.~\ref{fig:c2c4c6}, 
$c_6$ changes its sign at $T_c$. 
This means the peak position of $\chi_q$ moves left, 
which is corresponding to the change of $T_c$ as a function of $\mu_q$. 
$T_c(\mu_q/T=1)/T_c(\mu_q/T=0)$ in \cite{us02} is about $0.93$. 
Moreover, $\chi_q$ increases with higher orders of the expansion 
for $T \simlt T_c$ which confirms the existence of a peak. 
This suggests the presence of a critical endpoint in the $(T,\mu_q)$ plane.
On the other hand, $\chi_I$ in Fig.~\ref{fig:nsus} does not show any 
singular behavior. 
This is consistent with the sigma model prediction that only isosinglet 
degrees of freedom become massless at the critical endpoint.

\end{document}